\documentclass[aps,prb,onecolumn,amsmath,amssymb,floatfix,superscriptaddress]{revtex4-2}
\usepackage{graphicx}
\usepackage{bbm}
\usepackage[hidelinks]{hyperref}

\usepackage{color}
\setcitestyle{super}

\begin{document}
\title{Revised Enskog equation for hard rods}

\author{Vir B. Bulchandani}
\affiliation{Department of Physics, Princeton University, Princeton, New Jersey 08544, USA}
\affiliation{Institut f{\"u}r Theoretische Physik, Leibniz Universit{\"a}t Hannover, Appelstra{\ss}e 2, 30167 Hannover, Germany}

\begin{abstract}
We point out that Percus's collision integral for one-dimensional hard rods [J. K. Percus, Physics of Fluids 12, 1560–1563 (1969)] does not preserve the thermal equilibrium state in an external trapping potential. We derive a revised Enskog equation for hard rods and show that it preserves this thermal state exactly. In contrast to recent proposed kinetic equations for dynamics in integrability-breaking traps, both our kinetic equation and its thermal states are explicitly nonlocal in space. Our equation differs from earlier proposals at third order in spatial derivatives and we attribute this discrepancy to the choice of collision integral underlying our approach.
\end{abstract}
\maketitle

\section{Introduction}
Contemporary experiments\cite{kinoshita2006quantum,dipolar,atomchip,Malvania_2021} on systems of ultracold atoms have revealed a class of physical systems that do not thermalize efficiently under their own dynamics. Such experiments realize quasi-one-dimensional gases of particles with integrable two-body interactions (usually the Lieb-Liniger model of delta-interacting bosons on a line) and subject them to external trapping potentials that break microscopic integrability of the gas. The experimentally observed phenomenon of delayed thermalization in an integrability-breaking trap has been reproduced numerically in a variety of classical systems, including one-dimensional hard rods~\cite{Cao_2018,bagchi2023unusual,biagetti2023threestage}, the Toda lattice~\cite{DICINTIO2018249,dhar2019transport} and the rational Calogero model\cite{Bulchandani_2021}. 

Strikingly, the presence of thermalization in these systems appears to depend on the shape of the trapping potential; while there is evidence that for anharmonic trapping potentials, the system will eventually thermalize~\cite{bastianello2020thermalization,bagchi2023unusual,biagetti2023threestage}, there is also evidence that in harmonic trapping potentials, these systems can exhibit nonergodic behaviour at all numerically accessible times~\cite{Cao_2018,bagchi2023unusual}. This is surprising given that the only additional microscopic conservation law in a harmonic trap appears to be the centre-of-mass energy\cite{Cao_2018,bagchi2023unusual}. Although the short-time dynamics in a trap is widely believed to be captured by the generalized hydrodynamics (GHD) of integrable systems\cite{Castro_Alvaredo_2016,Bertini_2016,DoyonYoshimura_2017,bulchandani2017solvable,Cao_2018,Caux_2019,bastianello2020thermalization}, the validity of the latter description is questionable at long times because it implicitly assumes a perturbative treatment of integrability breaking that is not justified far from equilibrium~\cite{AGKL,intbreak}. Thus to address difficult questions concerning the presence or absence of thermalization at long times, a more systematic treatment of the trapping potential is desirable. 

In this paper, we propose such a treatment for systems of classical hard rods in an integrability-breaking trap. Such systems are appealing because they provide arguably the simplest nontrivial example of the phenomenology of interest; for example, the dynamics of hard rods between collisions is simply the dynamics of free particles in a trap. A single-particle kinetic equation for hard rods without a trap was first published by Percus~\cite{percus1969exact} in 1969; this equation was subsequently shown to yield exact hydrodynamic predictions at both the ballistic~\cite{boldrighini1983one} and diffusive~\cite{boldrighini1997one} scales. In a later paper\cite{percus1976equilibrium} in 1976, Percus derived the exact (grand canonical) thermal state of hard rods in a general confining external potential. However, the na{\"i}ve generalization of the 1969 kinetic equation to include a trapping potential contradicts the 1976 result describing trapped thermal states. This contradiction appears to have escaped notice until now and turns out to originate in the ``Enskog approximation'' to the equilibrium contact pair correlation function made in the earlier paper\cite{percus1969exact}. Below we calculate this pair correlation function exactly for trapped hard rods, and show that it leads to a small modification of Percus's kinetic equation that is nevertheless necessary to preserve the exact thermal state in a trap. Modifying Enskog's prescription in this way to yield a ``revised Enskog equation'' was previously advocated\cite{van1973modified} by Van Beijeren and Ernst as early as 1973 on general grounds. An analogous observation to ours was subsequently made~\cite{van1983equilibrium} for trapped three-dimensional hard spheres, but to the best of our knowledge the case of trapped hard rods has not been addressed. Motivation to revisit the latter comes from several puzzling recent numerical simulations~\cite{Cao_2018,bagchi2023unusual,biagetti2023threestage}.

The paper is structured as follows. We first summarize the state of knowledge on kinetic equations for trapped hard rods, and obtain a new characterization of the stationary states of the ballistic-scale kinetic equation. We then explain why existing kinetic theories are inconsistent with the known thermal state in a trap, and propose a revised Enskog equation that resolves this inconsistency. Finally, we perform a derivative expansion of the revised Enskog equation and show that it begins to differ from earlier proposals~\cite{Lebowitz68,spohn1982hydrodynamical,De_Nardis_2023} at third order in spatial derivatives. We argue that this discrepancy implies possible limitations on the validity of hydrodynamics as a model for long-time dynamics in integrability-breaking traps.

\section{Existing kinetic equations for hard rods}
\subsection{Kinetic equations without a trap}
In the absence of a trapping potential, Percus derived\cite{percus1969exact} a kinetic equation for hard rods by starting from the microscopic equation of motion
\begin{align}
\nonumber \partial_t \varrho_v + v \partial_x \varrho_v = &\int_{-\infty}^v dv' \, (v-v')[\varrho_v(x-a)\varrho_{v'}(x) - \varrho_v(x)\varrho_{v'}(x+a)] \\
\label{eq:BBGKY}
+ &\int_{v}^\infty dv' \, (v'-v)[\varrho_{v'}(x)\varrho_v(x+a) - \varrho_{v'}(x-a)\varrho_v(x)],
\end{align}
where
\begin{equation}
\varrho_v(x,t) = \sum_{i=1}^N \delta(x-x_i(t))\delta(v-v_i(t))
\end{equation}
denotes the microscopic phase-space distribution function. In order to do this, he used the Enskog form of Boltzmann's molecular chaos assumption
\begin{equation}
\label{eq:enskogapprox}
\langle \varrho_v(x)\varrho_{v'}(x+a) \rangle \approx \rho_v(x) \rho_{v'}(x+a) g(x+a/2),
\end{equation}
where the brackets indicate ensemble averaging with respect to an initial local generalized Gibbs state as is appropriate when considering the hydrodynamics of an integrable system\cite{Castro_Alvaredo_2016,Bertini_2016}, $\rho_v(x) = \langle \varrho_v(x)\rangle$ denotes the single-particle distribution function and $g(x+a/2)$ denotes the pair correlation function of a generalized Gibbs state with the same density of rods as at the midpoint of the rod centers at the instant of collision, given by\cite{tonks1936complete,bishop1974molecular}
\begin{equation}
\label{eq:approxPCF}
g(x+a/2) = \frac{1}{1-n(x+a/2)a},
\end{equation}
where $n(x)$ is the density of rods at position $x$. This yields the Enskog equation
\begin{align}
\nonumber \partial_t \rho_v + v \partial_x \rho_v  = &\int_{-\infty}^v dv' \, (v-v')[g(x-a/2)\rho_v(x-a)\rho_{v'}(x) - g(x+a/2) \rho_v(x)\rho_{v'}(x+a)] \\
\label{eq:origenskog}
+ &\int_{v}^\infty dv' \, (v'-v)[g(x+a/2)\rho_{v'}(x)\rho_v(x+a) - g(x-a/2)\rho_{v'}(x-a)\rho_v(x)].
\end{align}
We note that any translation-invariant state of the form
\begin{equation}
\label{eq:GGE}
\rho_v(x) = f(v),
\end{equation}
where $f : \mathbb{R} \to \mathbb{R}$ is a non-negative function, lies in the kernel of this collision integral for each $v'$; such states are trivially stationary under the hard-rod dynamics and define global generalized Gibbs ensembles for the untrapped hard rod gas\cite{boldrighini1983one,doyon2017dynamics}.

Truncating Eq. \eqref{eq:origenskog} at the ballistic (Euler) 
and diffusive (Navier-Stokes) scales yields kinetic equations that are known to be
exact in the appropriate scaling limits\cite{percus1969exact,boldrighini1983one,boldrighini1997one}, respectively
\begin{equation}
\label{eq:ballprevke}
\partial_t \rho_v + \partial_x \left(\frac{v - a j}{1-an}\rho_v\right) = 0
\end{equation}
at the ballistic scale, where
\begin{equation}
n = \int_{-\infty}^\infty dv' \, \rho_{v'}, \quad j = \int_{-\infty}^\infty dv' \, v'\rho_{v'},
\end{equation}
and
\begin{equation}
\label{eq:diffprevKE}
\partial_t \rho_v + \partial_x \left(\frac{v - a j}{1-an}\rho_v\right) = \frac{1}{2}a^2 \partial_x\left(\frac{1}{1-an} \int_{-\infty}^{\infty} dv' |v'-v|(\partial_x \rho_v \rho_{v'} - \rho_v \partial_x \rho_{v'})\right)
\end{equation}
if diffusive corrections are included. Both these equations preserve the global generalized Gibbs states Eq. \eqref{eq:GGE}. It is important to note that the mathematical derivation\cite{boldrighini1997one} of the diffusive-scale equation Eq. \eqref{eq:diffprevKE} requires a stronger notion of local equilibrium than is needed~\cite{boldrighini1983one} to derive Eq. \eqref{eq:ballprevke}. The former assumption does not always seem to be justified in practice~\cite{Cao_2018}. Thus there is a precise sense in which the diffusive-scale kinetic equation Eq. \eqref{eq:diffprevKE} is less accurate than the ballistic-scale kinetic equation Eq. \eqref{eq:ballprevke}.

\subsection{Kinetic theory in a trap}
The currently accepted\cite{DoyonYoshimura_2017,bastianello2020thermalization} diffusive-scale kinetic theory for hard rods in an external trapping potential $V(x)$ consists solely of adding a standard Boltzmann forcing term to the above equations. At the ballistic scale, this yields ($m=1$)
\begin{equation}
\label{eq:ballprevketrap}
\partial_t \rho_v + \partial_x \left(\frac{v - a j}{1-an}\rho_v\right) - V'(x) \partial_v \rho_v = 0,
\end{equation}
while including diffusive corrections, we have
\begin{equation}
\label{eq:diffprevketrap}
\partial_t \rho_v + \partial_x \left(\frac{v - a j}{1-an}\rho_v\right) - V'(x) \partial_v \rho_v = \frac{1}{2}a^2 \partial_x\left(\frac{1}{1-an} \int_{-\infty}^{\infty} dv' |v'-v|(\partial_x \rho_v \rho_{v'} - \rho_v \partial_x \rho_{v'})\right).
\end{equation}

In previous work, we both verified numerically that Eq. \eqref{eq:ballprevketrap} remained valid until a state-dependent ``time to chaos'' at which diffusive corrections became important, and pointed out that this equation could only be deduced from Eq. \eqref{eq:origenskog} if the trapping potential did not modify the Enskog approximation to the pair correlation function of the gas\cite{Cao_2018}. We will confirm below that revising the Enskog approximation yields no modifications to these equations at diffusive order~\cite{van1973modified} but to provide some additional motivation for this discussion, let us first consider the stationary states of Eqs. \eqref{eq:ballprevketrap} and \eqref{eq:diffprevketrap}. To understand stationary states of the ballistic scale dynamics Eq. \eqref{eq:ballprevketrap}, it is easiest to change variables to the ``free density''\cite{percus1969exact,doyon2017dynamics}
\begin{equation}
\theta_v = \rho_v/(1-an),
\end{equation}
to yield
\begin{equation}
\partial_t \theta_v + \left(\frac{v-aj}{1-an}\right) \partial_x \theta_v - V'(x) \partial_v \theta_v = 0,
\end{equation}
so that stationary states satisfy
\begin{equation}
\label{eq:stat}
\left(\frac{v-aj}{1-an}\right) \partial_x \theta_v = V'(x) \partial_v \theta_v.
\end{equation}
The possibility of infinitely many stationary states solving this equation was raised in previous work\cite{DoyonYoshimura_2017,Cao_2018}. Na{\"i}vely such states correspond to generalized Gibbs ensembles, but such states are not expected to survive the integrability-breaking effect of general trapping potentials. It has nevertheless been observed numerically that for hard rods in harmonic traps\cite{Cao_2018,bagchi2023unusual} and rational Calogero particles in arbitrary traps\cite{Bulchandani_2021}, non-Maxwellian stationary states persist for all accessible times.

Our first contribution below is an explicit construction of stationary states solving Eq. \eqref{eq:stat} for any differentiable trapping potential $V(x)$. To this end, let $F:\mathbb{R} \to \mathbb{R}$ be a differentiable, non-decreasing function with $F'(x) \geq 0$ for all $x \in \mathbb{R}$, such that there exists a solution $\epsilon_v(x)$ to the integral equation
\begin{equation}
\label{eq:inteq}
\epsilon_v = \frac{v^2}{2} + V(x) - \mu - a \int_{-\infty}^{\infty} dv'\,F(\epsilon_{v'}).
\end{equation}
Then $\theta_v = F'(\epsilon_v)$ solves Eq. \eqref{eq:stat}, since in this state, $j=0$ by evenness of $\epsilon_v$ in $v$ and one can show that
\begin{equation}
\partial_v \epsilon_v = v, \quad \partial_x \epsilon_v = (1-an)V'(x).
\end{equation}
Our construction, which is parameterized by the functional degree of freedom $F$, extends straightforwardly to the Euler-scale hydrodynamics of other quantum and classical integrable systems. For classical hard rods, we can compare this solution directly against exact results for the grand canonical thermal state in a trap\cite{percus1976equilibrium}. In the local density approximation (whose errors will be treated systematically below), the latter recovers Eq. \eqref{eq:inteq} with the specific choice
\begin{equation}
F(\epsilon) = - \frac{1}{\beta} e^{-\beta \epsilon}.
\end{equation}
The solutions corresponding to more general choices of $F$ can be viewed as maximizers of classical entropies that are not the usual Boltzmann-Gibbs entropy, which reflects the freedom to choose an entropy function in ballistic-scale hydrodynamics\cite{doyon2017dynamics}. The ballistic-scale hydrodynamics in a trap conserves an infinite family of such generalized entropies\cite{Cao_2018}.

It is interesting to consider whether these stationary states remain stationary when diffusive corrections are included. It was argued in previous work\cite{bastianello2020thermalization} that only thermal states in the above sense, satisfying $\theta_v = e^{-\beta \epsilon_v}$ and
\begin{equation}
\label{eq:LDATBAeq}
\epsilon_v = \frac{v^2}{2} + V(x) - \mu + \frac{a}{\beta} \int_{-\infty}^{\infty} dv'\,e^{-\beta \epsilon_{v'}}
\end{equation}
are stationary under the full dissipative evolution Eq. \eqref{eq:diffprevketrap}. We can see this directly by imposing vanishing of the dissipative part of the phase-space current in Eq. \eqref{eq:diffprevketrap} for general stationary states $\theta_v = F(\epsilon_v)$ of the ballistic scale equation, which yields the condition
\begin{equation}
\frac{F''(\epsilon_{v'})}{F'(\epsilon_{v'})} = \frac{F''(\epsilon_v)}{F'(\epsilon_v)}, \quad v' \neq v,
\end{equation}
on $F$. This implies that $F(\epsilon)$ is exponential in $\epsilon$, so that  the corresponding stationary state $\rho_v$ is always separable in $x$ and $v$ and therefore Maxwellian in $v$ by Eq. \eqref{eq:inteq}.

\section{Improved kinetic equation for hard rods}
\subsection{A contradiction and its resolution}
We now explain why the collision integral proposed by Percus\cite{percus1969exact} does not preserve thermal states in an external trapping potential. To our knowledge this observation has not been made before, although as we shall discuss below, it is consistent with van Beijeren and Ernst's critique\cite{van1973modified} of Enskog's prescription and van Beijeren's results for three-dimensional hard spheres\cite{van1983equilibrium}.

The single-particle distribution function for trapped thermal states of hard rods at inverse temperature $\beta$ and chemical potential $\mu$ is given by\cite{percus1976equilibrium}
\begin{equation}
\label{eq:GCEeq}
\rho_v(x) = \sqrt{\frac{\beta}{2\pi}}e^{-\beta v^2/2}n(x),
\end{equation}
where the local particle density $n(x)$ satisfies the nonlocal integral equation
\begin{equation}
\label{eq:percusnlie}
\log{n(x)} + \beta (V(x) - \mu) - \log{\sqrt{\frac{2\pi}{\beta}}} = \log{\left(1-\int_{x-a}^x dy \, n(y)\right)} - \int_x^{x+a} dy\, \frac{n(y)}{1-\int_{y-a}^y dz \, n(z)}.
\end{equation}
This result is exact in the grand canonical ensemble~\cite{percus1976equilibrium}, and by equivalence of ensembles should be exact in the canonical ensemble in the thermodynamic limit.

Let us start by interpreting Eqs. \eqref{eq:GCEeq} and \eqref{eq:percusnlie} for the exact thermal state in a trap. 
To relate these expressions to the thermodynamic-Bethe-ansatz-like equation Eq. \eqref{eq:LDATBAeq}, we define a nonlocally corrected version of the free density $\theta_v(x)$ in a trap, namely
\begin{equation}
\theta_v(x) = \frac{\rho_v(x)}{1-\int_{x-a}^x dy \, n(y)}.
\end{equation}
Writing $\theta_v(x) = e^{-\beta \epsilon_v(x)}$ as above, an exact and nonlocally corrected integral equation for $\epsilon_v$ is given by
\begin{equation}
\label{eq:TBAeq}
\epsilon_v(x) = \frac{v^2}{2} + V(x) - \mu + \frac{1}{\beta} \int_x^{x+a} dy\, \int_{-\infty}^{\infty} dv'\,e^{-\beta \epsilon_{v'}(y)}.
\end{equation}
This integral equation (or indeed Percus's original treatment\cite{percus1976equilibrium}) implies corrections to the local density approximation Eq. \eqref{eq:LDATBAeq} for trapped hard rods in thermal equilibrium\cite{bastianello2020thermalization,Kethepalli_2023} at all orders in the rod length $a$. We expect that such nonlocal terms are generically present for equilibrium states of interacting particles in integrability-breaking traps, 
and thus correct predictions\cite{bastianello2020thermalization} invoking the local density approximation, with the size of these corrections determined by the scattering length of the interactions.
 
We are now in a position to correct Eq. \eqref{eq:origenskog}. Let us suppose that the Enskog prescription Eq. \eqref{eq:approxPCF} remains valid in the presence of a trapping potential. This predicts the kinetic equation
\begin{align}
\nonumber \partial_t \rho_v + v \partial_x \rho_v - V'(x) \partial_v \rho_v = &\int_{-\infty}^v dv' \, (v-v')[g(x-a/2)\rho_v(x-a)\rho_{v'}(x) - g(x+a/2) \rho_v(x)\rho_{v'}(x+a)] \\
\label{eq:origtrapenskog}
+ &\int_{v}^\infty dv' \, (v'-v)[g(x+a/2)\rho_{v'}(x)\rho_v(x+a) - g(x-a/2)\rho_{v'}(x-a)\rho_v(x)].
\end{align}
However, upon substituting the exact equilibrium state Eq. \eqref{eq:GCEeq} into this equation, the left-hand side yields
\begin{equation}
\mathrm{L. \, H.\,S} = v\rho_v\left(\frac{n(x-a)}{1-\int_{x-a}^x dy \, n(y)} - \frac{n(x+a)}{1-\int_{x+a}^x dy \, n(y)}\right),
\end{equation}
while the right-hand side yields
\begin{equation}
\mathrm{R.\,H.\,S} = v\rho_v \left(\frac{n(x-a)}{1-an(x-a/2)} - \frac{n(x+a)}{1-an(x+a/2)}\right).
\end{equation}
The problem is now apparent: in an inhomogeneous trapping potential, Percus's proposed collision integral will not preserve the exact thermal state Eq. \eqref{eq:GCEeq}. To preserve this thermal state to all orders in the rod length $a$, we find that it suffices to replace the Enskog prescription Eq. \eqref{eq:approxPCF} by the approximation
\begin{equation}
\label{eq:modenskog}
\langle \varrho_v(x)\varrho_{v'}(x+a) \rangle \approx \rho_v(x) \rho_{v'}(x+a) g^{(2)}(x,x+a),
\end{equation}
where $g^{(2)}(x,x+a)$ denotes the exact nonuniform equilibrium contact pair correlation function of the hard rod gas. In Appendix \ref{sec:cpcf}, we show that for hard rods on an infinite line, in any potential that is confining as $|x| \to \infty$,
\begin{equation}
\label{eq:pcf}
g^{(2)}(x,x+a) = \frac{1}{1-\int_{x}^{x+a}dy \, n(y)}.
\end{equation}
We emphasize that Eq. \eqref{eq:modenskog} is a strict equality in thermal equilibrium within a trap, unlike the na{\"i}ve extrapolation from equilibrium without a trap\cite{tonks1936complete,bishop1974molecular} represented by Eq. \eqref{eq:approxPCF}. Moreover, this correction should persist even in the absence of a trap, which follows by considering arbitarily weak but still confining potentials $V(x) \to 0$.\\

We note that ``revising'' Enskog's prescription by using the exact nonuniform equilibrium pair correlation function at the instant of collision was previously advocated by van Beijeren and Ernst\cite{van1973modified,van1983equilibrium}, who argued that the resulting ``revised Enskog equation'' should be exact at sufficiently short times. For the specific case of trapped, three-dimensional hard spheres, van Beijeren further noted that revising Enskog's prescription was necessary in order to obtain thermal stationary states from kinetic theory~\cite{van1983equilibrium}. For trapped hard rods, the revised Enskog equation reads
\begin{align}
\nonumber \partial_t \rho_v + v \partial_x \rho_v - V'(x) \partial_v \rho_v = &\int_{-\infty}^v dv' \, (v-v')[g^{(2)}(x-a,x)\rho_v(x-a)\rho_{v'}(x) - g^{(2)}(x,x+a) \rho_v(x)\rho_{v'}(x+a)] \\
\label{eq:improvedke}
+ &\int_{v}^\infty dv' \, (v'-v)[g^{(2)}(x,x+a)\rho_{v'}(x)\rho_v(x+a) - g^{(2)}(x-a,x)\rho_{v'}(x-a)\rho_v(x)]
\end{align}
with $g^{(2)}(x,x+a)$ given by Eq. \eqref{eq:pcf}, and one can verify directly that this preserves the thermal state Eq. \eqref{eq:GCEeq}. The revised Enskog equation also respects  the appropriate microscopic conservation laws, namely conservation of total particle number and total energy in a generic trap and conservation of centre-of-mass energy in a harmonic trap.

We now characterize some of the stationary states of Eq. \eqref{eq:improvedke} for differentiable potentials $V(x)$. Specifically, we will show that any such potential for which arbitrary generalized Gibbs ensembles (GGEs) are stationary in Eq. \eqref{eq:improvedke} is uniform, and that all separable, time-reversal symmetric stationary states of Eq. \eqref{eq:improvedke} are thermal.

\subsection{Stationary states of the revised Enskog equation
}

\subsubsection{Stationary GGEs imply a uniform potential}
Generalized Gibbs states of the untrapped hard rod gas take the form of any spatially uniform profile
\begin{equation}
\rho_{v}(x) = f(v),
\end{equation}
for some non-negative function $f(v)$. Let us suppose that for some differentiable choice of trap $V(x)$, the kinetic equation Eq. \eqref{eq:improvedke} is stationary with respect to all such $\rho_v(x)$. Since the collision integral vanishes for any translation-invariant state, we deduce that
\begin{equation}
-V'(x) f'(v) = 0
\end{equation}
for all non-negative $f$, which implies that
\begin{equation}
V'(x) = 0,
\end{equation}
i.e. the potential is uniform.
\subsubsection{Separable stationary states are thermal}
Let us now suppose that Eq. \eqref{eq:improvedke} has a separable, time-reversal symmetric stationary state of the form 
\begin{equation}
\label{eq:sep}
\rho_v(x) = n(x) f(v)
\end{equation}
with $f(v)$ even in $v$. Without loss of generality we assume that $\int_{-\infty}^{\infty} dv \, f(v) = 1$, so that $n(x)$ is the local particle density. In general, the condition for stationarity can be written as
\begin{align}
\nonumber v \partial_x \rho_v - V'(x) \partial_v \rho_v = & g^{(2)}(x,x+a) \int_{-\infty}^\infty dv' \, (v'-v) \left(\mathbbm{1}_{v'<v} \rho_v(x)\rho_{v'}(x+a) +  \mathbbm{1}_{v'>v} \rho_{v'}(x)\rho_{v}(x+a)\right)\\
\label{eq:statcond}
-&g^{(2)}(x-a,x) \int_{-\infty}^\infty dv' \, (v'-v)\left(\mathbbm{1}_{v'<v}\rho_{v}(x-a)\rho_{v'}(x)+\mathbbm{1}_{v'>v}\rho_{v'}(x-a)\rho_{v}(x)\right).
\end{align}
For separable states Eq. \eqref{eq:sep}, this reduces to
\begin{align}
    \nonumber v f(v) \partial_x \log{n}(x) - V'(x) f'(v) = & \left(\frac{n(x+a)}{1-\int_{x}^{x+a}dy\, n(y)}
-\frac{n(x-a)}{1-\int_{x-a}^x dy \, n(y)}\right) \int_{-\infty}^\infty dv' \, (v'-v)f(v)f(v').
\end{align}
Using normalization and time-reversal symmetry of $f$, we can write this as
\begin{equation}
\partial_x \log{n(x)} + \frac{n(x+a)}{1-\int_{x}^{x+a}dy\, n(y)}
-\frac{n(x-a)}{1-\int_{x-a}^x dy \, n(y)} = V'(x) \frac{\partial_v \log{f(v)}}{v}.
\end{equation}
In order that the right hand side is independent of $v$, we require that
\begin{equation}
\partial_v \log f(v) = -\beta v
\end{equation}
for some constant $\beta$. By normalization of $f(v)$, this implies that
\begin{equation}
f(v) = \sqrt{\frac{\beta}{2\pi}}e^{-\beta v^2/2},
\end{equation}
i.e. $f(v)$ is Maxwellian. (Note that this constraint is only imposed if $V'(x) \neq 0$ for some $x$, so there is no contradiction with the previous result.) In particular, the density $n(x)$ must be related to the trapping potential via the equation
\begin{equation}
\partial_x \log{n(x)} + \beta V'(x) = \frac{n(x-a)}{1-\int_{x-a}^x dy \, n(y)} - \frac{n(x+a)}{1-\int_{x}^{x+a}dy\, n(y) },
\end{equation}
which is precisely the spatial derivative of Eq. \eqref{eq:percusnlie}. Thus $n(x)$ coincides with the thermal state in a trap up to an integration constant that we identify as the chemical potential.

\section{Derivative expansion of the revised Enskog equation}
In order to compare the predictions of our Eq. \eqref{eq:improvedke} against previous results on the higher-order hydrodynamics of hard rods\cite{Lebowitz68,spohn1982hydrodynamical,De_Nardis_2023}, which are usually stated as spatially local equations, we must develop an expansion in powers of the rod length $a$. At low orders, this expansion is identical to that of the Enskog equation Eq. \eqref{eq:origtrapenskog} because our proposed ``nonlocal'' correction to the pair correlation function
\begin{equation}
\label{eq:trapcorr}
g^{(2)}(x-a/2,x+a/2) - g(x) = \frac{a^3}{24}\frac{n''(x)}{(1-an(x))^2} + \mathcal{O}(a^5), \quad a \to 0,
\end{equation}
is only $\mathcal{O}(a^3)$ in the rod length, but at higher orders these nonlocal corrections will proliferate.

To obtain the derivative expansion of the revised Enskog equation, it is helpful to define the functional
\begin{equation}
G_{vv'}(x) = \frac{1}{1-\int_{x-a/2}^{x+a/2}dy \, n(y)} \rho_v(x-a/2)\rho_{v'}(x+a/2),
\end{equation}
which recovers the contact two-particle distribution function in equilibrium. In terms of this functional, and provided that $\rho_v(x)$ is smooth in $x$, the collision integral in Eq. \eqref{eq:improvedke} can be Taylor expanded as
\begin{align}
\nonumber I[\rho] &= \int_{-\infty}^v dv' (v-v')(G_{vv'}(x-a/2)-G_{vv'}(x+a/2)) +  \int_{v}^\infty dv' (v'-v)(G_{v'v}(x+a/2)-G_{v'v}(x-a/2)) \\
&= 2 \sum_{n=0}^\infty \frac{(a/2)^{2n+1}}{(2n+1)!} \partial_x^{2n+1} \left(\int_{-\infty}^v dv' (v'-v) G_{vv'}(x)+\int_{v}^\infty dv' (v'-v) G_{v'v}(x)\right).
\end{align}
To proceed further, we now expand (suppressing the argument of $g^{(2)}(x-a/2,x+a/2)$)
\begin{equation}
G_{vv'}(x) = g^{(2)}\sum_{m=0}^\infty \frac{(a/2)^m}{m!} \sum_{k=0}^m \begin{pmatrix}
m \\ k    
\end{pmatrix}
(-1)^k \rho_v^{(k)}(x) \rho_{v'}^{(m-k)}(x)
 \end{equation}
and note that these terms have alternating parity under interchange of velocities, to finally yield
\begin{align}
\nonumber I[\rho] = &2 \sum_{n,m=0}^\infty \frac{(a/2)^{2(n+m)+1}}{(2n+1)!(2m)!} \partial_x^{2n+1}  \left(g^{(2)}\int_{-\infty}^\infty dv' (v'-v) \sum_{k=0}^{2m}\begin{pmatrix} 2m \\ k \end{pmatrix} (-1)^k  \rho^{(2m-k)}_v(x) \rho^{(k)}_{v'}(x)\right) \\
\label{eq:fullderiv}
+ &2 \sum_{n,m=0}^\infty \frac{(a/2)^{2(n+m)+2}}{(2n+1)!(2m+1)!} \partial_x^{2n+1}  \left(g^{(2)} \int_{-\infty}^\infty dv' |v'-v| \sum_{k=0}^{2m+1}\begin{pmatrix} 2m+1 \\ k \end{pmatrix} (-1)^k \rho^{(2m+1-k)}_v(x) \rho^{(k)}_{v'}(x)\right).
\end{align}
Approximating $g^{(2)}(x-a/2,x+a/2)$ by $g(x)$ yields the full derivative expansion of Eq. \eqref{eq:origtrapenskog}.

Let us re-write the series expansion in Eq. \eqref{eq:fullderiv} as
\begin{equation}
\label{eq:org}
I[\rho] = \sum_{l=1}^\infty a^l I^{(l)}[\rho],
\end{equation}
where $I^{(l)}[\rho]$ can depend nontrivially on $a$ through $g^{(2)}$. It will be instructive to compute the first three terms in this series. For example, using Eq. \eqref{eq:trapcorr} and suppressing arguments, we find that
\begin{align}
aI^{(1)} = a\partial_x \left(\frac{j-nv}{1-an} \rho_v \right) + \frac{a^4}{24} \partial_x \left(\frac{n''(j-nv)}{(1-an)^2}\rho_v \right) + \mathcal{O}(a^6).
\end{align}
The first term is the usual velocity dressing for hard rods. The second term is new and arises from nonlocality of the pair correlation function $g^{(2)}(x,x+a)$. Similarly
\begin{equation}
a^2I^{(2)} = \frac{a^2}{2} \partial_x \left( \frac{1}{1-an} \int_{-\infty}^{\infty} dv' |v'-v|(\rho'_v \rho_{v'}-\rho_{v} \rho'_{v'})\right) + \mathcal{O}(a^5)
\end{equation}
recovers the usual diffusive term at leading order in $a$, with corrections arising from nonlocality at higher order in $a$ (here primes on $\rho$ denote spatial derivatives). Finally, we find that
\begin{equation}
a^3 I^{(3)} 
= \frac{a^3}{24} \left(\partial_x^3 \left( \frac{j-nv}{1-an}\rho_v\right) + 3 \partial_x \left(\frac{1}{1-an}\int_{-\infty}^\infty dv' (v'-v)(\rho''_v \rho_{v'}-2\rho'_{v} \rho'_{v'}+ \rho_v \rho_{v'}'') \right) \right) + \mathcal{O}(a^6).
\end{equation}
Combining these three expressions, we deduce that truncating the expansion Eq. \eqref{eq:org} at third order in spatial derivatives yields the kinetic equation
\begin{align}
\label{eq:dispketrap}
\nonumber &\partial_t \rho_v + \partial_x \left(\frac{v - a j}{1-an}\rho_v\right) - V'(x) \partial_v \rho_v = \frac{1}{2}a^2 \partial_x\left(\frac{1}{1-an} \int_{-\infty}^{\infty} dv' |v'-v|(\partial_x \rho_v \rho_{v'} - \rho_v \partial_x \rho_{v'})\right) \\
+ &\frac{a^3}{24}\left(\left(\partial_x \frac{an''}{1-an} + \partial_x^3\right)  \left(\frac{j-nv}{1-an}\rho_v \right) + 3 \partial_x \left(\frac{1}{1-an}\int_{-\infty}^\infty dv' (v'-v)(\rho''_v \rho_{v'}-2\rho'_{v} \rho'_{v'}+ \rho_v \rho_{v'}'') \right) \right).
\end{align}

The third-order term differs from a recent proposal for the dispersive corrections to generalized hydrodynamics\cite{De_Nardis_2023}, and indeed one can verify that the third-order contribution to Eq. \eqref{eq:dispketrap} deviates from this earlier prediction even at the level of linear response. This is perhaps surprising given that the latter was argued to recover known exact results for the hard-rod gas\cite{Lebowitz68,spohn1982hydrodynamical} near the Tonks-Girardeau limit of the Lieb-Liniger model\cite{De_Nardis_2023}. We expect that this discrepancy has its origin in the Enskog-Van Beijeren-Ernst form of Boltzmann's molecular chaos assumption Eq. \eqref{eq:modenskog}, which imposes a stringent notion of local equilibrium that does not necessarily hold in the earlier analytical derivations\cite{Lebowitz68,spohn1982hydrodynamical}.

\section{Conclusion}
Let us now review the context and the broader implications of our results. First, the context: the theory of generalized hydrodynamics~\cite{Castro_Alvaredo_2016,Bertini_2016} has emerged as the method of choice for simulating the finite-time dynamics of one-dimensional particles in trapping potentials that break integrability of their interactions~\cite{bulchandani2017solvable,Cao_2018,Caux_2019,atomchip}. In particular, the fact that the diffusive-scale equations in a trap predict thermalizing behaviour has been taken as evidence that thermalization is the ultimate fate of all such trapped integrable systems\cite{bastianello2020thermalization,biagetti2023threestage}. Numerical simulations of trapped hard rods are consistent with this scenario for anharmonic traps~\cite{bagchi2023unusual,biagetti2023threestage}, but exhibit a breakdown of both ergodicity and thermalization for harmonic traps on numerically accessible timescales~\cite{Cao_2018,bagchi2023unusual}.

Our paper challenges the current theoretical consensus in two main respects. First, the revised Enskog equation for trapped hard rods is explicitly nonlocal in space, in contrast to all earlier proposals for the kinetic theory of trapped integrable systems that we are aware of. Second, the subleading terms in the derivative expansion of the revised Enskog equation differ from the corresponding terms predicted by generalized hydrodynamics~\cite{De_Nardis_2023}, which are themselves consistent with earlier exact results~\cite{Lebowitz68,spohn1982hydrodynamical}. At the same time, it is difficult to imagine varying the collision integral of the revised Enskog equation while still preserving the thermal state within a trap; whether or not there exists a single kinetic equation that both preserves Percus's thermal state within a trap~\cite{percus1976equilibrium} and matches exact results without a trap~\cite{lebowitz1967kinetic,spohn1982hydrodynamical,De_Nardis_2023} remains to be seen. If no such collision integral exists, then it seems unlikely that the presence or absence of thermalization in an integrability-breaking trap can be deduced from generalized hydrodynamics alone.

To summarize, we believe that the single-particle kinetic equation Eq. \eqref{eq:improvedke} is accurate provided that its underlying assumption of local equilibrium Eq. \eqref{eq:modenskog} holds, and that this assumption can break down both in the absence of a trap~\cite{Cao_2018} and in the presence of a harmonic trap~\cite{Cao_2018,bagchi2023unusual}. The broader lesson from our analysis, which echoes earlier remarks in the literature~\cite{boldrighini1997one,Cao_2018} regarding the derivation of Eq. \eqref{eq:diffprevKE}, is that any single-particle kinetic theory within a trap that accounts for diffusion must assume some notion of local equilibrium, and will only yield accurate predictions at long times (for example, regarding the presence or absence of thermalization) if this notion of local equilibrium is preserved under time evolution within the trap in question. From this viewpoint, the question of why some trapping potentials appear to sustain microscopic ergodicity better than other trapping potentials\cite{bagchi2023unusual} lies beyond hydrodynamics, and demands a more microscopic explanation. An important open question is whether any of various proposed microscopic mechanisms\cite{Cao_2018,bastianello2020thermalization,intbreak,biagetti2023threestage} for thermalization within a trap can explain these discrepancies.

\section{Acknowledgements}
V.B.B. was supported by a fellowship at the Princeton Center for Theoretical Science during part of the completion of this work and thanks X. Cao, A. Dhar, F. Essler, D.A. Huse, M. Kulkarni and J.E. Moore for helpful discussions.

\appendix
\section{The exact contact pair correlation function}
\label{sec:cpcf}
In this Appendix, we derive the exact contact pair correlation function for trapped hard rods in thermal equilibrium using the method of Percus\cite{percus1976equilibrium}, who did not explicitly consider this quantity although it can be deduced from his results. To motivate Percus's method, which requires working in the grand canonical ensemble, it is instructive to first attempt the same derivation in the canonical ensemble.
\subsection{Canonical ensemble}
Consider $N$ hard rods on an infinite line in an external trapping potential $V(x)$ in thermal equilibrium at inverse temperature $\beta$. The canonical partition function can be written as 
\begin{equation}
Z = Z_{k,N} Z_{N}
\end{equation}
where the ``kinetic'' partition function
\begin{equation}
\label{eq:defkin}
Z_{k,N} = \prod_{i=1}^N  \int_{-\infty}^\infty dp_i\, e^{-\beta p_i^2/2} = \left(\frac{2\pi}{\beta}\right)^{N/2}
\end{equation}
and the ``configurational'' partition function
\begin{align}
Z_{N} &= \int_{-\infty}^\infty dx_1 \int_{x_1+a}^{\infty} dx_2 \ldots \int_{x_{N-1}+a}^{\infty} dx_N \, e^{-\beta \sum_{j=1}^N V(x_j)} \\
&= \int_{-\infty}^\infty dx_N \int_{-\infty}^{x_N-a} dx_{N-1} \ldots \int_{-\infty}^{x_{2}-a} dx_1 \, e^{-\beta \sum_{j=1}^N V(x_j)} 
\end{align}
can be written in two different ways depending on whether the rods are enumerated from left to right or right to left respectively. Then the one-particle density can be written as
\begin{equation}
n(x) = e^{-\beta V(x)} \frac{1}{Z_N} \sum_{M=0}^{N-1} Z_{M}(-\infty,x)Z_{N-1-M}(x,\infty)
\end{equation}
where the sum is over all possible arrangements of $M$ rods to the left and $N-1-M$ rods to the right of the rod centred at $x$ and the one-sided partition functions
\begin{equation}
Z_{M}(-\infty,x) = \int_{-\infty}^{x-a} dx_M \int_{-\infty}^{x_M-a} dx_{M-1} \ldots \int_{-\infty}^{x_{2}-a} dx_1 \, e^{-\beta \sum_{j=1}^M V(x_j)}
\end{equation}
and
\begin{equation}
Z_{M}(x,\infty) = \int_{x+a}^{\infty} dx_1 \int_{x_1+a}^{\infty} dx_2 \ldots \int_{x_{M-1}+a}^{\infty} dx_M \, e^{-\beta \sum_{j=1}^M V(x_j)}
\end{equation}
weight these arrangements appropriately, with the convention that $Z_{0}(-\infty,x) = Z_{0}(x,\infty) = 1$. Similarly, the contact two-particle distribution function can be written as
\begin{equation}
n^{(2)}(x,x+a) = e^{-\beta V(x)}e^{-\beta V(x+a)} \frac{1}{Z_N} \sum_{M=0}^{N-2} Z_{M}(-\infty,x)Z_{N-2-M}(x+a,\infty).
\end{equation}
In general, these expressions are difficult to extract predictions from as $N \to \infty$, but their form is suggestive of multiplying power series with coefficients $Z_M$. This in turn suggests that dramatic simplifications could occur in the grand canonical ensemble, which indeed turns out to be the case.

\subsection{Grand canonical ensemble}
Let us now consider the grand canonical partition for hard rods on an infinite line, following Percus\cite{percus1976equilibrium}. This 
can be written as
\begin{equation}
\Xi = \sum_{N=0}^\infty z^N Z_{k,N} Z_N.
\end{equation}
where the fugacity $z = e^{\beta \mu}$. Introducing an effective fugacity
\begin{equation}
\tilde{z} = z \sqrt{\frac{2\pi}{\beta}},
\end{equation}
we can write this as
\begin{equation}
\Xi = \sum_{N=0}^\infty \tilde{z}^N Z_N.
\end{equation}
Then, introducing the one-sided grand canonical partition functions
\begin{equation}
\Xi(-\infty,x) = \sum_{N=0}^{\infty} \tilde{z}^N Z_{N}(-\infty,x)
\end{equation}
and
\begin{equation}
\Xi(x,\infty) = \sum_{N=0}^{\infty} \tilde{z}^N Z_{N}(x,\infty),
\end{equation}
we can write the one-particle density as\cite{percus1976equilibrium}
\begin{equation}
n(x) = \frac{1}{\Xi} \tilde{z} e^{-\beta V(x)} \Xi(-\infty,x)\Xi(x,\infty)
\end{equation}
and the contact two-particle distribution function as
\begin{equation}
n^{(2)}(x,x+a) = \frac{1}{\Xi} \tilde{z}^2 e^{-\beta V(x)} e^{-\beta V(x+a)}\Xi(-\infty,x)\Xi(x+a,\infty).
\end{equation}
This implies a simple formula for the contact pair correlation function, namely
\begin{equation}
g^{(2)}(x,x+a) = \frac{n^{(2)}(x,x+a)}{n(x)n(x+a)} = \frac{\Xi}{\Xi(-\infty,x+a)\Xi(x,\infty)}.
\end{equation}
Then, using the formula\cite{percus1976equilibrium}
\begin{equation}
\Xi(-\infty,x+a)\Xi(x,\infty) = \Xi\left(1-\int_x^{x+a} dy \, n(y)\right),
\end{equation}
we deduce that the exact contact pair correlation function for trapped hard rods in thermal equilibrium is given by
\begin{equation}
\label{eq:exactpcf}
g^{(2)}(x,x+a) = \frac{1}{1-\int_x^{x+a} dy \, n(y)}.
\end{equation}

We emphasize that this ``equation of state'' for the pair correlation function holds for even for generalized-Gibbs-type ensembles with an arbitrary probability distribution in momentum space (although these are not strictly stationary except in the limit that $V(x)=0$). For such ensembles, the configurational partition function is unchanged while the kinetic partition function is modified to
\begin{equation}
Z_{k,N} = \prod_{i=1}^N  \int_{-\infty}^\infty dp_i\, e^{-f(p_i)} = \left(\int_{-\infty}^\infty dp \, e^{-f(p)}\right)^{N}
\end{equation}
where the function $f$ uniquely determines the generalized Gibbs ensemble in question. In the treatment above, this change only modifies the effective fugacity $\tilde{z}$ and therefore does not affect Eq. \eqref{eq:exactpcf}.

\bibliography{bibl}
\end{document}